\documentclass{article}
\usepackage[english]{babel}

\usepackage[a4paper,top=2cm,bottom=2.5cm,left=3cm,right=3cm,marginparwidth=1.75cm]{geometry}

\usepackage{amsmath}
\usepackage{graphicx}
\usepackage{amssymb}
\usepackage[colorlinks=false, linkcolor=black, citecolor=black]{hyperref}
\usepackage{tikz}
\usetikzlibrary{tqft,calc}

\title{ \textbf{Bosonic String Dynamics}}
\author{Henrique Legoinha \\ 
{\small h.legoinha@cern.ch} \\ 
{\small Instituto Superior Técnico, Lisbon}}
\date{January 15, 2024}

\begin{document}
\maketitle

\begin{abstract}
String Theory is a hot topic of physics and mathematics. For the former, it stands as a huge sandbox where the formulation of difficult problems can be simplified and their hard computations carried out. For the latter, it stands as a direct contribution upstream, going from physics to maths, rather than the usual other way around. These notes provide a comprehensive review of the bosonic string action, the entering point of String Theory, and also covers the resulting equations of motion and the effective mass of open and closed strings. 
\end{abstract}

\clearpage

\hrule
\vspace{-4mm}
\renewcommand*\contentsname{}
\tableofcontents
\vspace{6mm}
\hrule
\vspace{10mm}

\section{Introduction}

\begin{figure}[h!]
   \centering
    \begin{tikzpicture}[scale=1]

    \draw[ thick, blue] (0.75,1,-1) arc (0:-150:0.5);
    \draw[ thick, blue] (0.75,1,-1) arc (180:120:0.5);
    \fill[blue] (1,1.42,-1) circle (1.pt);
    \fill[blue] (-0.18,0.75,-1) circle (1.pt);
    \draw[ thick, blue] (4,1.3,0) ellipse (1. and 0.5);

    \end{tikzpicture}
    \caption{Heuristic sketch of what an open and closed string look like.}
    \label{ocstring}
\end{figure}
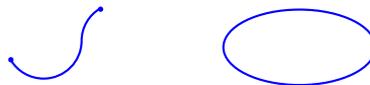

The core idea behind String Theory is to consider 1-dimensional objects, the so called strings, rather than 0-dimensional objects, the point particles, as the fundamental subject of study. Thus, String Theory stands as the simplest modification of quantum field theory and general relativity that can be achieved. The strings have no sort of internal structure and, as portrayed in Fig.~\ref{ocstring}, either feature an open or closed topology. Open strings and closed strings are not objects governed by different theories but rather distinct states inside the same theory. In fact, theories of open strings necessarily require closed strings since as soon as interactions are allowed open strings can become closed strings. 

When studying open strings it quickly follows that the theory would be very incomplete if the dynamics of further higher dimensional objects, the so called branes, are left out. While strings have, for sure, a central focus, it is important to keep in mind that branes also play a pivotal role in String Theory. Hence, String Theory is actually the theory of any d-dimensional object living inside some (d+1)-dimensional space.

There are several good motivations to learn and further invest in String Theory. From an ambitious perspective, String Theory aims to deliver a theory of everything and, in this matter, it is the leading proposal. It happens that open strings describe Yang-Mills theories while closed strings describe Gravity \cite{IST}. Since open strings can close and vice versa, Gravity and Yang-Mills theories are automatically dynamically related under a single theory. In contrast, other concurrent theories for quantum gravity do not offer such a unified picture, thus being not so gorgeous. From a practical perspective, String Theory provides a huge framework
to study strongly coupled field theories through a quite radical manifestation of the principle of holography, the duality between Anti de Sitter space and Conformal Field Theory, AdS/CFT \cite{maldacena}. This means that a weakly coupled String Theory living on an AdS is equivalent to some strongly coupled conformal field theory. Hence, perturbative string computations can be carried out to give insights into the associated hardly solvable field theory. A tentative and highly wanted application of this duality is to the low energy region of Quantum Chromodynamics, were no perturbative computation holds. Still, QCD is not at all a conformal field theory and so some strategies to make the duality work must be considered \cite{Erdmenger_2008}. From a diplomatic perspective String Theory is at the center of a prosperous interplay between mathematicians and physicists. String Theory serves as a giant sandbox for applying modern mathematical concepts to physics and, in turn, the intuition of string theorists has catalyzed entirely new developments within the realm of pure mathematics.
For these reasons it is indeed worth learning String Theory, at least to some extent.

\section{String Dynamics}

A string is embedded in some $(1+p)$-dimensional \textit{target space} where it must always be in motion through time. Doing so it sweeps out a $(1+1)$-dimensional \textit{worldsheet}, similarly to a point particle, which sweeps a $(1+0)$-dimensional \textit{worldline}. Hence, the geometric interpretation of the least action principle demands that such surface stands for an extreme, usually a minimum, and so strings evolve through paths which minimize the traced area. For this reason we want to parameterize the worldsheet and so two coordinates are employed, one with a timelike nature, $\tau$, and one with a spacelike nature, $\sigma$. Therefore, $X^\mu(\tau,\sigma)$, where $\mu=0,...,p-1$, is a map giving the the target space coordinates of any point on the worldsheet. In these notes we take the target space to be some $p$-dimensional Minkowsky space, with a metric given by $\eta = \text{diag}(-++++...)$.

\subsection{The Nambu-Goto action }

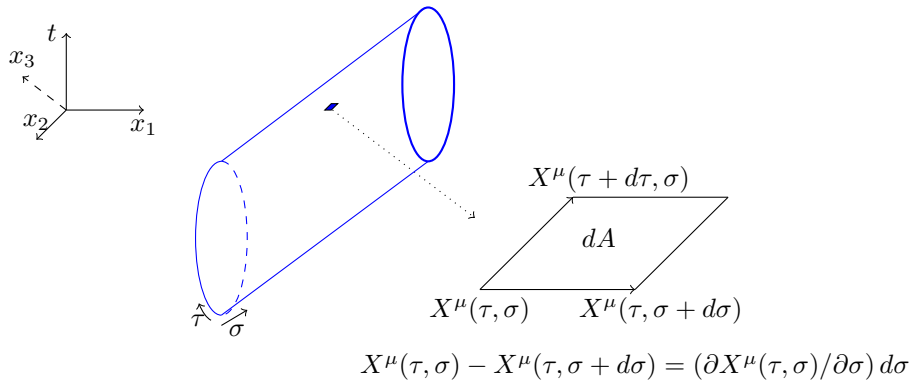
\begin{figure}[h!]
   \centering
\begin{tikzpicture}[scale=0.68]{

    \draw[->] (-3,1,0) -- (-1.5,1,0) node[below]{$x_1$};
    \draw[->] (-3,1,0) -- (-3,2.5,0) node[left]{$t$};
    \draw[->] (-3,1,0) -- (-3,1,1.5) node[above]{$x_2$};
    \draw[dashed, ->] (-3,1,0) -- (-4,1.5,-0.4)node[above]{$x_3$};

  \draw[fill=blue] (2,1) -- (2.125,1.) -- (2.25,1.125) -- (2.125,1.125) -- cycle;

    \draw[dotted, ->] (2.15,1) -- (4.9,-1.1);

  \coordinate (A) at (5,-2.5);

    \draw[->] (A) -- ++(3,0);
    \draw[->] (A) -- ++(1.8,1.8);
    \draw (6.8,-0.7) -- ++(3,0);
    \draw (8,-2.5) -- ++(1.8,1.8);

  \draw (8.5,-2.9) node {$X^\mu(\tau,\sigma+d\sigma)$};
  \draw (5,-2.9) node {$X^\mu(\tau,\sigma)$};
  \draw (7.5,-0.35) node {$X^\mu(\tau+d\tau,\sigma)$};

 \draw (7.3,-1.5) node {$dA$};

   \draw (8,-4) node {$ X^\mu(\tau,\sigma)-X^\mu(\tau,\sigma+d\sigma)= (\partial X^\mu(\tau,\sigma)/\partial \sigma) \, d\sigma$};
  
  \draw[->] (-0.2,-3.10) arc (-80:-50:-0.5 and 1.5);
  \draw[blue] (0,-3) -- (4,0.);
  \draw[blue] (0,0) -- (3.95,3);
  \draw[blue] (0,-3) arc (270:90:0.5 and 1.5);
    \draw[dashed, blue] (0,0) arc (90:-90:0.5 and 1.5);
  \draw[thick, blue] (4,1.5) ellipse (0.5 and 1.5);

  \draw (0.3,-3.3) node {$\sigma$};
  \draw[->] (0.,-3.2) -- (.5,-2.9);
  \draw (-0.45,-3.1) node {$\tau$};

}
\end{tikzpicture}
    
    \caption{Sketch of a closed string worldsheet and its infinitesimal of area }
    \label{AreaAction}
\end{figure}

In order to write the action governing the dynamics of a string we need to find the worldsheet area in terms of the spacetime coordinates, $X^\mu$, whose dependence on $\tau$ and $\sigma$ is implicit from now on. Being an embedded surface on spacetime the worldsheet has an induced metric, $\gamma_{ab}$, given by the pull-back of the Minkowski metric,
\begin{align}
    \gamma_{ab} = \frac{\partial X^\mu}{\partial \sigma^a} \frac{\partial X^\nu}{\partial \sigma^b} \eta_{\mu\nu}
\end{align}
where $a=0,1$ and $\sigma^a=(\tau,\sigma)$ groups the two worldsheet coordinates. Hence, the string action in natural units, $\hbar=c=1$, reads
\begin{align}
    \label{NGa}
    S_{NG}&=-T\int d\tau d\sigma \sqrt{-\text{det}(\gamma)}  
\end{align}
where $T=1/2\pi\alpha'$ is the string tension. This action, Eq.~(\ref{NGa}), is known as the Nambu-Goto action and its derivation can be easily followed with the visual help of Fig.~\ref{AreaAction}, which tells us that we need to find the expression for the area of the sketched parallelogram. Hence, recalling the formula for the area of a parallelogram of sides $a$ and $b$, $A= (|a|^2|b|^2-(a\cdot b)^2)^{1/2}$, we see that
\begin{align}
    \label{areaF}
       dA &= d\tau d\sigma \sqrt{ \Bigg| \frac{\partial X}{\partial \sigma} \Bigg|^2 \Bigg| \frac{\partial X}{\partial \tau} \Bigg|^2 - \Bigg(  \frac{\partial X}{\partial \sigma} \cdot\frac{\partial X}{\partial \tau}\Bigg)^2 } \nonumber \\
       &= d\tau d\sigma \sqrt{-\text{det}(\gamma)} 
\end{align}
and so the integral in the Nambu-Goto action is indeed evaluating the area of the worldsheet.

Inspecting Eq.~(\ref{NGa}) carefully we find that there are two types of symmetries in the Nambu-Goto action. One is Poincaré invariance, which is a global symmetry from the perspective of the worldsheet since when $X^\mu \rightarrow \Lambda^\mu_\nu X^\nu + c^\mu$ the Lorentz transformations, $\Lambda^\mu _\nu$, and the translations, $c^\mu$, have no dependence on the worldsheet coordinates. Then there is reparameterization invariance, which is a gauge symmetry reflecting the fact that the worldsheet coordinates have no physical meaning at all. In fact, changing the worldsheet coordinates can not alter the area that a string sweeps and so the action is blind to this choice, as it should.

\clearpage

\subsection{The Polyakov action }

The square root in the Nambu-Goto action will lead to difficulties when working out the theory in the path integral formalism. However, written as in Eq.~(\ref{NGa}) there is an hidden gauge symmetry, local conformal invariance, associated to the fact that the worldsheet is 2-dimensional. This can be exploited to push the induced metric out of the square root. Conformal transformations can be best described as some angle preserving re-scaling of the metric,
\begin{align}
\label{ConformalT}
    \gamma_{ab} \longrightarrow h_{ab} = \Omega^2 \gamma_{ab}
\end{align}
where the conformal factor, $\Omega^2$, has some implicit dependence on the worldsheet coordinates. Two metrics related by a conformal transformation must be associated with the same physical state. Hence, writing $\gamma_{ab} = \Omega^{-2} h_{ab}$ we can find a more interesting expression for the string action,
\begin{align}
    S_{NG} \longrightarrow &-T \int d\tau d\sigma \sqrt{-\text{det}(h)} \; \sqrt{\Omega^{-4}} \;  \nonumber \\
    = &-T \int d\tau d\sigma \sqrt{-\text{det}(h)} \; \Omega^{-2}.
\end{align}
Therefore, if we ask for the conformal factor to be $\Omega^{-2} = \gamma\,/2 = \gamma^a_a/2 = h^{ab}\gamma_{ab}/2$, with the one half factor being introduced here just for future convenience, we have successfully shifted the induced metric out of the square root,  
\begin{align}
\label{PolyA}
    S_{P} &= -\frac{T}{2} \int d\tau d\sigma \sqrt{-det(h)} \; h^{ab} \; \gamma_{ab} \; \nonumber \\
        &= -\frac{T}{2} \int d\tau d\sigma \sqrt{-det(h)} \; h^{ab} \; \partial_aX^\mu \partial_bX^\nu,
\end{align}
at the cost of introducing some auxiliary tensor field $h_{ab}$ which now takes the role of the worldsheet metric. This new expression for the string action, Eq.~(\ref{PolyA}), is known as Polyakov action and it makes the local conformal symmetry visible. For instance, it is straightforward to show that under a transformation like the one in Eq.~(\ref{ConformalT}) the action remains unaltered since $\sqrt{-\text{det}(h)}\rightarrow \Omega^2 \sqrt{-\text{det}(h)}$ and $h^{ab} \rightarrow \Omega^{-2} h^{ab}$.

There is a striking coincidence between the Polyakov action governing string dynamics and the Einstein-Hilbert action governing gravity. In fact, the Polyakov action for string propagation in $\mathbb{R}^{1,d-1}$ appears to be describing (d+1)-scalar fields coupled to 2-dimensional gravity. Hence, in String Theory the spacetime coordinates $X^\mu$ are promoted to dynamical scalar fields living in a worldsheet where a dynamical tensor field, the metric $h_{ab}$, also lives. 

The Polyakov version of the string action can still be simplified. Since the worldsheet metric tensor $h$ is symmetric, it features three independent components, each function of the two coordinates $\tau$ and $\sigma$, it is possible to set any two of them to a value of our interest. Then, taking the reparameterization invariance of the action together with its conformal symmetry we can demand the metric to become locally conformal flat,
\begin{align}
    h_{ab} = \Omega^2 \eta'_{ab},
\end{align}
where $\eta'= \text{diag}(-+)$. This is known as conformal gauge. However, as discussed above, the conformal factor drops out when inserted in the Polyakov action and so it drastically simplifies to
\begin{align}
\label{freeF}
    S_{P} \longrightarrow &-\frac{T}{2} \int d\tau d\sigma \sqrt{-\text{det}(\eta)}\; \sqrt{\Omega^4}\; \Omega^{-2}\eta^{ab}\; \gamma_{ab} \nonumber \\
    =& -\frac{T}{2} \int d\tau d\sigma \sqrt{-\text{det}(\eta)} \; \eta^{ab}\gamma_{ab} \nonumber \\
    =& -\frac{T}{2} \int d\tau d\sigma \; \eta^{ab}\gamma_{ab}  \nonumber \\
    =& -\frac{T}{2} \int d\tau d\sigma \; \partial_a X^\mu \partial^a X_\mu,
\end{align}
which is nothing but the action for a theory of free massless scalar fields. Thus, the fundamental object in String Theory are really just fields. Nonetheless, all this simplicity is not dropping out of the sky. There are constrains that shall be discussed next, while deriving the equations of motion.

\subsection{Equations of motion}

Owning to gauge symmetry the Polyakov action is cast from Eq.~(\ref{PolyA}) to Eq.~(\ref{freeF}) since we have freedom to chose an extremely simple metric, $h_{ab} \longrightarrow \eta'_{ab}$. However, the straightforwardness of the last expression might conceal the fact that there must also be an equation of motion for the metric tensor in addition to an equation of motion for the scalar field. Using the Euler-Lagrange equation it is almost effortless to find both fields' equations of motion. Lets start by writing the scalar field one, 
\begin{align}
\label{waveM}
   \frac{\partial \mathcal{L}}{\partial X^\mu} &= \partial_a \bigg( \frac{\partial \mathcal{L}}{\partial (\partial_a X^\mu)} \bigg) \nonumber \\
    0 &= \partial_a \bigg( -\frac{T}{2} \, \frac{\partial }{\partial (\partial_a X^\mu)}\Big( \partial_b X^\nu \, \partial^b X_\nu \Big) \bigg) \nonumber \\
     0 &= \partial_a ( \partial^a X^\mu + \partial^a X^\mu ) \nonumber \\
    0 &= \partial_a \partial^a X^\mu,
\end{align}
which is, as expected, just a free wave equation. Lets now turn our attention to the equation of motion of the metric,
\begin{align}
\label{metricM}
  \partial_c \bigg( \frac{\partial \mathcal{L}}{ \partial (\partial_c \eta'^{ab})} \bigg) &= \frac{\partial \mathcal{L}}{\partial \eta'^{ab} } \nonumber \\
0 &= -\frac{T}{2} \, \frac{\partial}{\partial \eta'^{ab}} \Big( \eta'^{cd}\,\partial_{c}X_\mu \,\partial_{d}X^\mu \Big) \nonumber \\
0 &= \partial_a X^\mu \, \partial_b X_\mu
\end{align}

We used the Euler-Lagrange equation to arrive at the above equations of motion for each respective field entering in the Polyakov action. However, the Euler-Lagrange equation follows from demanding that the variation of the action vanishes and thus ensuring that the principle of least action is satisfied. Hence, in the case of the metric tensor, we can identify Eq.~(\ref{metricM}) with the stress-energy tensor of the theory, $T_{ab}$, since that is the quantity that follows from varying the action with respect to the metric, $T_{ab}=\delta S/\delta\eta'^{ab}$. Therefore we see that it vanishes and, writing it more explicitly, we find
\begin{align}
    \label{cons1}
    T_{00} = T_{11} = \dot{X}^2 + X'^2  = 0 \\
    \label{cons2}
    T_{01} = T_{10} = \dot{X} \cdot X' = 0
\end{align}
with $\dot{X} = \partial_0 X$ and $X' = \partial_1 X$, meaning that the equation of motion governing the behavior of a string is a simple free wave equation, Eq.~(\ref{waveM}), subject to the two constrains that follow from $T_{ab}=0$, Eq.~(\ref{cons1}) and Eq.~(\ref{cons2}), known as Viraroso constrains. Furthermore, since the stress-energy tensor is conserved, $\partial_a T^{ab}$ = 0, as can be readily concluded by looking at Eq.~(\ref{waveM}), the Viraroso constrains can be imposed just once on an initial time. 

Some insight about the behaviour of strings can be obtain by inspecting the first constrain, Eq.~(\ref{cons1}), were we learn that the parameterization of the worldsheet must be such that lines of constant $\sigma$ are orthogonal to lines of constant $\tau$. This implies that strings can only move perpendicularly to the direction in which they are stretched. Also for this reason strings can never feature longitudinal oscillations, which agrees with their lack of internal structure. 

Up to this point it is irrelevant if the string is open or closed. In fact, according to the formulation so far, the dynamics of a worldsheet point $X(\tau,\sigma)$ is governed by local physics, meaning that it has no clue if it belongs to an open or closed string. Nonetheless, there are constrains arising from differences in the parameter space of open and closed strings that the respective string solution to the equation of motion, Eq.~(\ref{waveM}), must also respect. Namely, the spacelike coordinate $\sigma$ for open strings lives in the range $[0,\pi]$ while for closed strings it lives in the two folded bigger range, $[0,2\pi[$. Logically, for a closed string it must be required that 
\begin{align}
\label{periodic}
    X^\mu(\tau,\sigma)=X^\mu(\tau,\sigma+2\pi).
\end{align} 
In turn, the lack of this constrain for open strings leads to a very important emergence of two other constrains that allow to satisfy the principle of least action. These are rather subtle, and appear solely to ensure that the derivations based on the Euler-Lagrange equation achieved above hold and so, for this reason, their discussion has been postponed until now. It happens that when varying of the action, $\delta S $, an unusual term appears in the usual step of integration by parts and, for the case of open strings, it requires special attention to put it down and thus ensure that $\delta S =0$. Accordingly, we see that
\begin{align}
    \delta S &= -\frac{T}{2} \int_{\tau_i}^{\tau_f} \int_{\sigma=0}^{\sigma=l_s} \, d\tau \, d\sigma\; \Big\{ 2\partial_a X^\mu \partial^a \delta X_\mu + \Big[\partial_a X^\mu \, \partial_b X_\mu\Big]  \delta \eta'^{ab} \Big\}  \nonumber \\
    &= -\frac{T}{2} \int_{\tau_i}^{\tau_f} \, \int_{0}^{l_s} \, d\tau \, d\sigma \Big\{ -2 \Big[\partial^a\partial_aX^\mu\Big] \delta X_\mu + \Big[ T_{ab} \Big]\delta \eta'^{ab} + 2\partial^a(\delta X_\mu\partial_a X^\mu) \, \Big\},
\end{align}
where $l_s=\pi$ for open strings and $l_s=2\pi$ for closed strings. The terms inside the square brackets are demanded to be zero, thus yielding the Euler-Lagrange equations of motion for the respective fields, but this can only be the case if the last term vanishes as well. After integrating it we find
\begin{align}
\label{missingt}
 \int_{\tau_i}^{\tau_f}\int_{0}^{l_s} d\tau d\sigma  \; 
 \partial^a(\delta X_\mu \partial_a X^\mu) \, =& - \, \Bigg[ \int_{0}^{l_s} d\sigma \, \dot{X}^\mu \, \delta X_\mu \Bigg]_{\tau_i}^{\tau_f} \, +\, \Bigg[ \int_{\tau_i}^{\tau_f} d\tau \, X'^\mu \,\delta X_\mu \Bigg]_{0}^{l_s}   \\
 =& \, \Bigg[ \int_{\tau_i}^{\tau_f} d\tau \, X'^\mu \,\delta X_\mu \Bigg]_{0}^{l_s}
\end{align}
where the second term in Eq.~(\ref{missingt}) is known to be zero since the principle of least action is based on the vanishing of $\delta X_\mu$  at times $\tau_i$ and $\tau_f$. Still, the second term is unusual and only simplifies naturally for closed strings where there is the periodic constrain, Eq.~(\ref{periodic}), ensuring $X^\mu(\tau,0)=X^\mu(\tau,2\pi)$ so that the integrand vanishes. For open strings, to make this term go away, there are two possible boundary conditions that can be demanded, namely, Neumann boundary conditions, 
\begin{align}
    X'^\mu=0 \quad\quad \text{at $\sigma=0,\pi$ }
\end{align}
and Dirichlet boundary conditions,
\begin{align}
    \delta X^\mu = 0 \quad\quad \text{at $\sigma=0,\pi$}.
\end{align}
Each of these leads to different phenomena. For instance, Neumann boundary conditions have no restriction on $\delta X^\mu$, allowing the end-points of a string to move freely in space. In direct contrast, Dirichlet boundary conditions fixes $\delta X^\mu$, thus forcing the string end-points to be permanently fixed at some spacetime points, $X^\mu(\tau,0)=c^\mu$ and $X^\mu(\tau,\pi)=c'^\mu$.

\subsection{D\textit{p}-brane action}

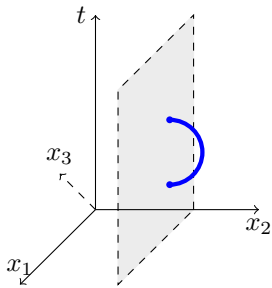
\begin{figure}[h!]
   \centering
    \begin{tikzpicture}[scale=0.86]

    \filldraw[gray!15] (1.5,0,0) -- (1.5,3,0) -- (1.5,3,3) -- (1.5,0,3) -- cycle;
    \draw[dashed] (1.5,0,0) -- (1.5,3,0) -- (1.5,3,3) -- (1.5,0,3) -- cycle;

    \draw[ultra thick, blue] (0.75,1,-1) arc (90:-90:0.5);

    \fill[blue] (0.75,1.,-1) circle (1.5pt);
    \fill[blue] (0.75,0,-1) circle (1.5pt);
    \draw[->] (0,0,0) -- (2.5,0,0) node[below]{$x_2$};
    \draw[->] (0,0,0) -- (0,3.,0) node[left]{$t$};
    \draw[->] (0,0,0) -- (0,0,3.) node[above]{$x_1$};
    \draw[dashed, ->] (0,0,0) -- (-0.7,0.4,-0.4)node[above]{$x_3$};

    \end{tikzpicture}
    \caption{Sketch of a D$2$-brane embedded in a 4-dimensional target space with a string attached.}
    \label{memb}
\end{figure}

It is possible to mix both Neumann and Dirichlet boundary conditions, for instance,
\begin{align}
\label{res1}
    X'^i = 0~\wedge~X^j = c^j &\quad\quad \text{at $\sigma=0,\pi$ } \quad\quad \text{for $i=0,...,p$ ~ and ~ $j=p+1,...,(d-1)$ } 
\end{align}
with $d$ the dimension of the target space where the string is embedded. Doing so the restrictions given by Eq.~(\ref{res1}) force both end-points of the string to live inside a same $(p+1)$-dimensional hypersurface, known as D$p$-brane, where D stands for Dirichlet and $p$ is the number of spatial dimensions of the hypersurface. In fact, the case where Neumann boundary conditions are applied in all direction should actually be pictured as a D$p$-brane extending everywhere. In Fig.~\ref{memb} there is a sketch of a string whose end-points are attached to the same D$2$-brane. 

String Theory is not just a theory of strings, it is actually a theory of D$p$-branes. For instance, a D$0$-brane traces a worldline throughout spacetime and thus can be identified as a point particle. Similarly, a D$1$-brane draws a worldsheet throughout spacetime and so it is to be identified as a string. In this line of thinking any higher dimensional object dynamics must also be accounted for. In fact, a theory of gravity can not contain rigid objects. For this reason there must be an action for them as well. It is easy to find it by just generalizing the Nambu-Goto action, Eq.~(\ref{NGa}), to p-dimensions,
\begin{align}
    \label{Branea}
    S_{Dp}&=-T_p\int d\xi^{p+1} \sqrt{-\text{det}(\gamma_{p})}  
\end{align}
with $T_p$ the tension of the D$p$-brane and $\gamma_p$ the induced metric on the $(p+1)$-dimensional surface traced by the D$p$-brane. This is known as the Dirac action.

It might appear that D$1$-branes have caught more our attention than their higher dimensional counterparts. Among the reasons why, it is because that after quantization a discrete spectrum where each state is associated with a particle is present. However, in huge contrast, the spectrum that arises after quantization of a higher dimensional D$p$-branes is continuous and so it is expected to describe multi-particle states \cite{Tong}.  

\section{Solutions on the Light-Cone}

We are going to solve the string equation of motion, Eq.~(\ref{waveM}), switching to lightcone coordinates on the worldsheet,
\begin{align}
    \sigma^{\pm} &= \tau \pm \sigma \nonumber \\
    \partial_{\pm} &= \frac{1}{2}\Big(\frac{\partial}{\partial \tau} \pm \frac{\partial}{\partial \sigma} \Big) \nonumber
\end{align}
so that the equation of motion becomes
\begin{align}
    \partial_a \partial^a X^\mu  = 0\; \longrightarrow& \;\; \partial_+ \partial_- X^\mu= 0
\end{align}
and the Viraroso constrains become,
\begin{align}
\label{viralight}
    T_{ab} = 0 \; \longrightarrow& \;\; (\partial_+)^2 = (\partial_-)^2 = 0 .
\end{align}

A general solution to the equation of motion can then be found through separation of variables,
\begin{align}
    X^\mu(\tau,\sigma) = X_L^\mu(\sigma^+) + X_R^\mu(\sigma^-)
\end{align}
with $X_L$ and $X_R$, which describe left and right moving waves, given by an expansion in terms of Fourier modes,
\begin{align}
    X_L^\mu(\sigma^+) &= \frac{1}{2}x^\mu + A\alpha'p^\mu \sigma^+ + i\sqrt{\frac{\alpha'}{2}}\sum_{n\neq0}\frac{1}{n} \Tilde{\alpha}_n^\mu e^{-in\sigma^+} \\
    X_R^\mu(\sigma^-) &= \frac{1}{2}x^\mu + A\alpha'p^\mu \sigma^- + i\sqrt{\frac{\alpha'}{2}}\sum_{n\neq0}\frac{1}{n} \alpha_n^\mu e^{-in\sigma^-}
\end{align}
where $A=1$ for open strings and $A=1/2$ for closed strings. For closed strings $X_L$ and $X_R$ do not individually satisfy the periodicity condition, only their sum does. Reality of the solution implies that $\Tilde{\alpha}^\mu_n = (\Tilde{\alpha}^\mu_{-n})^*$ and similarly for $\alpha^\mu_n$. These describe the amplitudes of the oscillations on the string and the remaining different coefficients ensure that $x^\mu$ and $p^\mu$ can be associated with the position and momentum of the center of mass of the string. This can be checked by finding the conserved Noether current arising from the spacetime translation symmetry, $X^\mu \rightarrow X^\mu(\tau,\sigma) + c^\mu$. The best way to do so is by employing the Noether trick \cite{Tong2}, where the translation is momentarily promoted to a local transformation, $c^\mu \rightarrow c^\mu(\tau,\sigma)$
\begin{align}
    \mathcal{L}= -\frac{T}{2} \partial_a X^\mu \partial^a X_\mu \; \longrightarrow& \; -\frac{T}{2} \partial_a X^\mu \partial^a X_\mu + 2 \partial^a X^\mu \partial_a c_\mu \nonumber \\
     =& \; \mathcal{L} \, \underbrace{- T (\partial_a c_\mu)\partial^a X^\mu}_\text{ $\delta\mathcal{L}$ }.
\end{align}

Therefore the change in the action is 
\begin{align}
    \delta S &= \int d\tau d\sigma \; \delta \mathcal{L} = \nonumber \\
            &= \int d\tau d\sigma \Big\{-T (\partial^a c_\mu) \partial^a X^\mu \Big\}                  \nonumber \\
            &= \int d\tau d\sigma \Big\{\partial^a(-Tc_\mu \partial_a X^\mu ) + c_\mu \partial^a(T\partial_aX^\mu) \Big\}          \nonumber \\
            &= \int d\tau d\sigma \; c^\mu \, \partial^a \, j_a^\mu 
\end{align}
with $j_a^\mu = T \partial_a X^\mu$. According to Eq.~(\ref{waveM}) $\partial^a j_a^\mu = 0$ and therefore we may identify the conserved 
Noether current to be $j_a^\mu$ and the conserved charge associated to it, $Q^\mu$, to be
\begin{align}
    Q^\mu = \int_0^{l_s} d\sigma j_0^\mu = T \int_0^{l_s} d\sigma \dot{X}^\mu.
\end{align}
Substituting for $ X^\mu(\tau,\sigma) = X_L^\mu(\sigma^+) + X_R^\mu(\sigma^-)$ and considering the adequate boundary conditions for the string in question it follows that $p^\mu$ is the conserved charge. For example, for the case of closed strings 
\begin{align}
    Q^\mu &= \int_0^{2\pi} d\sigma \; j_0^\mu\; =  T \int_0^{2\pi} d\sigma\frac{\partial}{\partial \tau} \Big\{ X_L^\mu(\sigma^+) + X_R^\mu(\sigma^-) \Big\} \nonumber \\
    &= \frac{1}{2\pi\alpha'}  \int_0^{2\pi} d\sigma \Bigg\{ \alpha'p^\mu + \sqrt{\alpha'/2}\sum_{n\neq0}\Big( \Tilde{\alpha}^\mu_n e^{-in\sigma^+} - \alpha^\mu_n e^{-in\sigma^+}\Big) \Bigg\} \nonumber \\
    &= \frac{1}{2\pi\alpha'} \Bigg\{ 2\pi\alpha'p^\mu + \sqrt{\alpha'/2}\sum_{n\neq0}\Big( \Tilde{\alpha}^\mu_n e^{-in\sigma}\Big[^{2\pi}_0 - \alpha^\mu_n e^{+in\sigma}\Big[^{2\pi}_0 \Big)e^{-in\tau} \Bigg\} \nonumber \\
    &= p^\mu
\end{align}
The case of open strings is no more complicated as soon as the Neumann boundary conditions imposed, in turn implying that $\alpha^\mu_n=\Tilde{\alpha}^\mu_n$.

So far the Viraroso constrains have not yet been applied to the equation of motion for $X^\mu$. Doing so will reveal the mass spectra of strings as function of the oscillation modes and is the subject of the following section.

\subsection{String mass}

Open and closed strings belong to the same theory but have, in their own right, different expressions for the mass. Lets start by inspecting what happens in the case of closed strings. Imposing the Viraroso constrains, Eq.~(\ref{viralight}), results in a condition for the oscillation modes, namely
\begin{align}
    (\partial_-X)^2 &= \frac{\alpha'}{2} \sum_{m,k} \, \alpha_m \cdot \alpha_k \, e^{-i(m+k)\sigma^-} \nonumber \\
     &= \frac{\alpha'}{2} \sum_{m,n} \, \alpha_m \cdot \alpha_{n-m} \, e^{-in\sigma^-} \nonumber \\
     &= \alpha' \sum_n L_n e^{-in\sigma^-} = 0
\end{align}
where $L_n $ is the sum of oscillator modes given by 
\begin{align}
    L_n = \sum_n \frac{1}{2} \alpha_{n-m} \cdot \alpha_m
\end{align}
and where the 0-th mode is 
\begin{align}
    \alpha_0^\mu = \sqrt{\frac{\alpha'}{2}} p^\mu
\end{align}
A similar derivation follow for the left-moving modes, yielding analogous expressions. Together they translate the Viraroso constrains into oscillation modes constrains. However, classically, imposing validity of the Virasoro constraints is equivalent to requiring $L_n = 0 \, \forall n$ \cite{IntroSS}. Therefore the constrain arising from $L_0 = 0$ has a very important meaning since it includes the square of the string momentum, which in turn is known to equal the square of the mass, $p^\mu p_\mu=-M^2$. Hence we arrive at an expression for an effective mass of the closed string in terms of the excited oscillation modes,
\begin{align}
    L_0 &= \, \frac{1}{2}\sum_m \alpha_{0-m}\cdot \alpha_m \nonumber \\
        &= \, \frac{1}{2}\alpha_0\cdot\alpha_0 \, + \, \sum_{m>0}\alpha_{m}\cdot\alpha_{-m}  \nonumber \\
        &= \frac{\alpha'}{4}p^2+ \, \sum_{m>0}\alpha_{m}\cdot\alpha_{-m}   \nonumber \\
       &= -\frac{\alpha'M^2}{4}+ \, \sum_{m>0}\alpha_{m}\cdot\alpha_{-m}  \; = \, 0  \nonumber \\
        &\Rightarrow M^2 = \frac{4}{\alpha'} \,\sum_{m>0}\alpha_{m}\cdot\alpha_{-m}
\end{align}
Using instead the left-moving modes would yield the exact same expression for the mass since $\alpha_0=\Tilde{\alpha}_0$. This goes by the name of \textit{level matching} and could not have been any other way, otherwise the theory would be ill-defined.

We now turn our attention to open strings. Lets consider that it lives in between two D$p$-branes, thus having the following Dirichlet boundary conditions,
\begin{align}
    X^I(\tau,0) = c^I~\wedge~X^I(\tau,\pi)=d^I
\end{align}
which in turn imply that $p^I=0$ and $\alpha_n^I = -\Tilde{\alpha}_n^I$. This way we see that there is only one set of oscillators and, in term of the Fourier mode expansion we have
\begin{align}
    X^I &= (X_L^I(\sigma^+) + X_R^I(\sigma^-))  \nonumber \\
    &= x^I - i \sqrt{\frac{\alpha'}{2}}\sum_{n\neq0} \frac{1}{n}\alpha^I_n \, e^{-in\sigma^+} + i \sqrt{\frac{\alpha'}{2}}\sum_{n\neq0} \frac{1}{n}\alpha^I_n \, e^{-in\sigma^+} \nonumber \\
    &= x^I + i \sqrt{\frac{\alpha'}{2}}\sum_{n\neq0}\Big\{ \frac{1}{n}\alpha_n^{I}  \Big( e^{in\sigma} - e^{-in\sigma}\Big) e^{-in\tau} \Big\} \nonumber \\
    &= c^I \frac{(d^I-c^I)\sigma}{\pi}+ i \sqrt{\frac{\alpha'}{2}}\sum_{n\neq0}\Big\{ \frac{1}{n}\alpha_n^{I}  \Big( e^{in\sigma} - e^{-in\sigma}\Big) e^{-in\tau} \Big\}
\end{align}
Then, imposing the Viraroso constrains we arrive at
\begin{align}
    (\partial_+X)^2 = \alpha'^2p^2 + \frac{|\Vec{d}-\Vec{c}|^2}{4\pi^2} + \text{(oscillator modes)} = 0,
\end{align}
from where we once again learn that the string mass is a function of the excited modes,
\begin{align}
\label{masso}
    M^2= \frac{|\Vec{d}-\Vec{c}|^2}{4\alpha'^2\pi^2} + \text{(oscillator modes)}
\end{align}
There is, however, an extra term in comparison to the mass expression of the closed strings. This new term means that there are contributions to the mass of a classical open string if it is stretched. Hence, the larger the stretching, the larger the mass.

There is one more interesting scenario to consider, the case where there are $N$ branes and all are stacked at the same place. In this situation there are $N^2$ possible ways of attaching a string end-points to a brane. To distinguish each case it is normal to use the \textit{Chan-Paton factor}, a label $m,n = 1,...,N$ that is given to the string ends according to the $n$-th brane where they sit. Then, each string in this scenario will have a mass expression given by Eq.~(\ref{masso}), thus standing for a total of $N^2$ different particles.

\section{Final Regards}

We have discussed the very basic foundations of String Theory as well as some key results over the course of the last pages. First, the geometric interpretation of the principle of least action lead us to the Nambu-Goto action which next was cast into the version of the Polyakov action. The latter is much preferred since it better encapsulates what String Theory looks like, a theory of gravity. From here, we obtained and discussed the several constrains that solutions to the equations of motion must obey for both string types. Lastly, exploring them we were able to write an expression for the mass of open and closed strings.

There is, however, a huge elephant in the room. Bosonic String Theory just by itself is doomed since it contains tachyonic degrees of freedom. This problem can be readily solved by imposing super-symmetry. To achieve it we simple add fermionic counterparts to the action, which due to its familiar form in Eq.~(\ref{freeF}) is not very hard to guess how such terms will look like. 

The next logical step is to quantize the theory, which can either be  done canonically or through the path integral formalism. The later has the advantage of allowing a formulation of non-pertubative phenomena as well, a situation that not even String Theory is free of. In fact, most of the higher-dimensional objects predicted by the theory are hardly pertubative. When quantizing using path integrals it will be necessary to correct the redundant integration over physically equivalent states using the Faddeev-Popov method. 

A future study direction in the context of String Theory is conformal field theory or topology. In the end, the main idea of a pertubative computation in String Theory goes like the familiar version in quantum field theory. To achieve some computation in the framework of String Theory it will be be necessary to sum over the possible topologies of the worldsheet and, in parallel, results coming from conformal field theory are needed to make sense of the associated diagrams. Good knowledge of conformal field theories are also needed to establish the AdS/CFT duality and proceed with computations in this correspondence. 

\bibliographystyle{ieeetr}
\bibliography{sample}

\end{document}